\documentclass{article}

\usepackage[utf8]{inputenc}
\usepackage[USenglish]{babel}
\usepackage{enumitem}
\usepackage{tikz}
\usepackage{verbatim}
\usepackage{graphicx}
\usepackage[a4paper,margin=2cm]{geometry}% A4 for a US project?
\usepackage{lineno}

\usepackage{amssymb}% http://ctan.org/pkg/amssymb
\usepackage{amsmath}
\usepackage{pifont}% http://ctan.org/pkg/pifont
\usepackage[backend=biber,defernumbers=true,sorting=none,giveninits=true,natbib=true,style=numeric-comp]{biblatex}

\renewbibmacro{in:}{}
\begin{document}
\title{Machine Learning in High Energy Physics Community White Paper}

\maketitle

\setlength\parindent{0pt}
\textbf{Abstract:}
Machine learning has been applied to several problems in particle physics research, beginning with applications to high-level physics analysis in the 1990s and 2000s, followed by an explosion of applications in particle and event identification and reconstruction in the 2010s.
In this document we discuss promising future research and development areas for machine learning in particle physics. We detail a roadmap for their implementation, software and hardware resource requirements, collaborative initiatives with the data science community, academia and industry, and training the particle physics community in data science.
The main objective of the document is to connect and motivate these areas of research and development with the physics drivers of the High-Luminosity Large Hadron Collider and future neutrino experiments and identify the resource needs for their implementation.
Additionally we identify areas where collaboration with external communities will be of great benefit.

\vskip 1cm
\textbf{Editors}: Sergei Gleyzer$^{30}$, Paul Seyfert$^{13}$, Steven Schramm$^{32}$\\
\newline
\textbf{Contributors}:
Kim Albertsson$^{1}$,
Piero Altoe$^{2}$,
Dustin Anderson$^{3}$,
John Anderson$^{4}$,
Michael Andrews$^{5}$,
Juan Pedro Araque Espinosa$^{6}$,
Adam Aurisano$^{7}$
% ...
\bigskip
\par {\footnotesize $^{1}$ Lulea University of Technology}
\par {\footnotesize $^{2}$ NVidia}
\par {\footnotesize $^{3}$ California Institute of Technology}
\par {\footnotesize $^{4}$ Google}
\par {\footnotesize $^{5}$ Carnegie Mellon University}
\par {\footnotesize $^{6}$ LIP Lisboa}
\par {\footnotesize $^{7}$ University of Cincinnati}
\par {\footnotesize $^{8}$ Universita e INFN, Padova}
\par {\footnotesize $^{9}$ University of London}
\par {\footnotesize $^{10}$ Lawrence Berkeley National Laboratory}
\par {\footnotesize $^{11}$ Universita e INFN, Bologna}
\par {\footnotesize $^{12}$ Brown University}
\par {\footnotesize $^{13}$ CERN}
\par {\footnotesize $^{14}$ Argonne National Laboratory}
\par {\footnotesize $^{15}$ CPPM Aix Marseille Univ CNRS/IN2P3}
\par {\footnotesize $^{16}$ Universitaet Freiburg}
\par {\footnotesize $^{17}$ New York University}
\par {\footnotesize $^{18}$ Deutsches Elektronen-Synchrotron}
\par {\footnotesize $^{19}$ Duke University}
\par {\footnotesize $^{20}$ SISSA Trieste Italy}
\par {\footnotesize $^{21}$ Fermi National Accelerator Laboratory}
\par {\footnotesize $^{22}$ RWTH Aachen University}
\par {\footnotesize $^{23}$ Universit{\"a}t Z{\"u}rich}
\par {\footnotesize $^{24}$ University of Texas at Arlington}
\par {\footnotesize $^{25}$ Southern Methodist University}
\par {\footnotesize $^{26}$ Ecole Polytechnique Federale de Lausanne}
\par {\footnotesize $^{27}$ University of Manchester}
\par {\footnotesize $^{28}$ University of Cambridge}
\par {\footnotesize $^{29}$ Karlsruher Institut f{\"u}r Technologie}
\par {\footnotesize $^{30}$ University of Florida}
\par {\footnotesize $^{31}$ LPNHE, Sorbonne Universit\'{e} et Universit\'{e} Paris Diderot, CNRS/IN2P3, Paris}
\par {\footnotesize $^{32}$ Universit{\'e} de Gen{\`e}ve}
\par {\footnotesize $^{33}$ Louisiana Tech University}
\par {\footnotesize $^{34}$ Purdue University}
\par {\footnotesize $^{35}$ University of Illinois at Urbana-Champaign}
\par {\footnotesize $^{36}$ Max Planck Institut f{\"u}r Physik}
\par {\footnotesize $^{37}$ SLAC National Accelerator Laboratory}
\par {\footnotesize $^{38}$ SoundCloud}
\par {\footnotesize $^{39}$ University of Milan}
\par {\footnotesize $^{40}$ Intel}
\par {\footnotesize $^{41}$ Brookhaven National Laboratory}
\par {\footnotesize $^{42}$ University of Bristol}
\par {\footnotesize $^{43}$ Russian Academy of Sciences}
\par {\footnotesize $^{44}$ Cornell University}
\par {\footnotesize $^{45}$ University of Notre Dame}
\par {\footnotesize $^{46}$ University of Melbourne}
\par {\footnotesize $^{47}$ University of California Berkeley}
\par {\footnotesize $^{48}$ Universita \& INFN, Milano Bicocca}
\par {\footnotesize $^{49}$ Vrije Universiteit Brussel}
\par {\footnotesize $^{50}$ University of Amsterdam and Radboud University Nijmegen}
\par {\footnotesize $^{51}$ Yale University}
\par {\footnotesize $^{52}$ University of Alabama}
\par {\footnotesize $^{53}$ University of Massachusetts}
\par {\footnotesize $^{54}$ Princeton University}
\par {\footnotesize $^{55}$ Florida State University}
\par {\footnotesize $^{56}$ Indiana University}
\par {\footnotesize $^{57}$ College of William and Mary}
\par {\footnotesize $^{58}$ University of California, Santa Cruz}
\par {\footnotesize $^{59}$ Rice University}
\par {\footnotesize $^{60}$ Universite de Paris Sud 11}
\par {\footnotesize $^{61}$ University of Oklahoma}
\par {\footnotesize $^{62}$ Masaryk University}
\par {\footnotesize $^{63}$ University of Glasgow}
\par {\footnotesize $^{64}$ Radboud Universiteit Nijmegen}
\par {\footnotesize $^{65}$ Altair Engineering}
\par {\footnotesize $^{66}$ Ohio Supercomputer Center}
\par {\footnotesize $^{67}$ Yandex School of Data Analysis}
\par {\footnotesize $^{68}$ Technical University of Kosice}
\par {\footnotesize $^{69}$ Gangneung-Wonju National University}
\par {\footnotesize $^{70}$ University of Rochester}
\par {\footnotesize $^{71}$ University of Barcelona}
\par {\footnotesize $^{72}$ University of Chicago}
\par {\footnotesize $^{73}$ University of Washington}
\par {\footnotesize $^{74}$ Massachusetts Institute of Technology}
\par {\footnotesize $^{75}$ Chinese Academy of Sciences}
\par {\footnotesize $^{76}$ OProject and University of Antioquia}

\clearpage

\tableofcontents
\clearpage

\section{Preface}
\label{sec:preface}
To outline the challenges in computing that high-energy physics will face over the next years and strategies to approach them, the HEP software foundation has organised a Community White Paper (CWP)~\cite{mainCWP}. In addition to the main document, several more detailed documents were worked out by different working groups. The present document focusses on the topic of machine learning.
The goals are to define the tasks at the energy and intensity frontier that can be addressed during the next decade by research and development of machine learning applications.\\ % In this context, the interaction of the high-energy physics and data science communities is discussed, prioritized areas of machine learning R\&D are proposed, machine learning software and hardware platforms and resources are analyzed and a roadmap of future sustainable development is presented.

Machine learning in particle physics is evolving fast, while the contents of this community white paper were mainly compiled during community meetings in spring 2017 that took place at several workshops on machine learning in high-energy physics: S2I2 and \cite{DSatHEP2017,IML2017,ACAT2017,HSF2017}.
The contents of this document thus reflect the state of the art at these events and does not attempt to take later developments into account.

\section{Introduction}
\label{sec:introduction}
% ---------------------------------------------------
% CHAPTER SUMMARY MESSAGE:  We want to identify the science drivers and the primary goals
% and key areas for using ML
% ---------------------------------------------------

%As particle physics enters the post-Higgs boson discovery era, one of the main objectives is

One of the main objectives of particle physics in the post-Higgs boson discovery era is to exploit the full physics potential of both the Large Hadron Collider (LHC) and its upgrade, the high luminosity LHC (HL-LHC), in addition to present and future neutrino experiments.
The HL-LHC will deliver an integrated luminosity that is 20 times larger than the present LHC dataset, bringing quantitatively and qualitatively new challenges due to event size, data volume, and complexity. The physics reach of the experiments will be limited by the physics performance of algorithms and computational resources. Machine learning (ML) applied to particle physics promises to provide improvements in both of these areas.\\

Incorporating machine learning in particle physics workflows will require significant research and development over the next five years. Areas where significant improvements are needed include:
\begin{itemize}
 \item \textbf{Physics performance} of reconstruction and analysis algorithms;
 \item \textbf{Execution time} of computationally expensive parts of event simulation, pattern recognition, and calibration;
 \item \textbf{Realtime implementation} of machine learning algorithms;
 \item \textbf{Reduction of the data footprint} with data compression, placement and access.
\end{itemize}
\noindent

\subsection{Motivation}
The experimental high-energy physics (HEP) program revolves around two main objectives that go hand in hand: probing the Standard Model (SM) with increasing precision and searching for new particles associated with physics beyond the SM. Both tasks require the identification of rare signals in immense backgrounds. Substantially increased levels of pile-up collisions from additional protons in the bunch at the HL-LHC will make this a significant challenge.\\

%Machine learning community benefits from rapid development and experimentation. Modern machine-learning tools are themselves undergoing a process of rapid evolution. Noteworthy are the recent machine-learning tools originating from industry. The HEP community can benefit from these advancements by developing flexible software and workflows. A major challenge is how to engage the ML community and maximally benefit from these developments. Common to the HEP and ML communities is the desire to interpret machine learning models~\cite{interpretability}.

Machine learning algorithms are already the state-of-the-art in event and particle identification, energy estimation and pile-up suppression applications in HEP. Despite their present advantage, machine-learning algorithms still have significant room for improvement in their exploitation of the full potential of the dataset.

% an open area of research expected to require a substantial R\&D  effort. %ML developments for HEP are likely to have applications for other scientific domains which are similarly exposed to large amounts of data.

\subsection{Brief Overview of Machine Learning Algorithms in HEP}

This section provides a brief introduction to the most important machine learning algorithms in HEP, introducing key vocabulary (in \textit{italic}).\\

%Specific application areas of machine learning in HEP are detailed in Chapter ~\ref{sec:applications}.

Machine learning methods are designed to exploit large datasets in order to reduce complexity and find new features in data. The current most frequently used machine learning algorithms in HEP are Boosted Decision Trees (BDTs) and Neural Networks (NN).\\

%\textcolor{red}{DR: PROBABLY NEED DIAGRAMS}
Typically, variables relevant to the physics problem are selected and a machine learning \textit{model} is \textit{trained} for \textit{classification} or \textit{regression} using signal and background events (or \textit{instances}).
Training the model is the most human- and CPU-time consuming step, while the application, the so called \textit{inference} stage, is relatively inexpensive.
BDTs and NNs are typically used to classify particles and events.
They are also used for regression, where a continuous function is learned, for example to obtain the best estimate of a particle's energy based on the measurements from multiple detectors.\\

Neural Networks have been used in HEP for some time; however, improvements in training algorithms and computing power have in the last decade led to the so-called deep learning
revolution, which has had a significant impact on HEP. Deep learning is particularly promising when there is a large amount of data and features, as well as symmetries and complex non-linear dependencies between inputs and outputs.\\

There are different types of DNN used in HEP: fully-connected (FCN), convolutional (CNN) and recurrent (RNN). Additionally, neural networks are used in the context of Generative Models, where a Neural Network is trained to reproduce the multidimensional distribution of the training instances set. Variational AutoEncoders (VAE) and more recent Generative Adversarial Networks (GAN) are two examples of such generative models used in HEP.\\

A large set of machine learning algorithms is devoted to time series analysis and prediction. They are in general not relevant for HEP data analysis where events are independent from each other. However, there is more and more interest in these algorithms for Data Quality, Computing and Accelerator Infrastructure monitoring, as well as those physics processes and event reconstruction tasks where time is an important dimension.

\subsection{Structure of the Document}

%With many machine learning tools available and a long standing tradition of homegrown HEP software,

Applications of machine learning algorithms motivated by HEP drivers are detailed in Section~\ref{sec:applications}, while Section~\ref{sec:collaboration} focuses on outreach and collaboration with the machine learning community. Section~\ref{sec:software} focuses on the machine learning software in HEP and discusses the interplay between internally and externally developed machine learning tools. Recent progress in machine learning was made possible in part by emergence of suitable hardware for training complex models, thus in Section~\ref{sec:resources} the resource requirements of training and applying machine learning algorithms in HEP are discussed. Section~\ref{sec:training} discusses strategies for training the HEP community in machine learning.  Finally, Section~\ref{sec:roadmap} presents the roadmap for the near future.\\

\section{Machine Learning Applications and R\&D}
\label{sec:applications}
% ---------------------------------------------------------
% CHAPTER SUMMARY MESSAGE: < Explain using science drivers the R&D needed>
% ---------------------------------------------------------

% Contributor information SHOULD BE SAVED, but can be left as a comment for now
%Contributors: Mark Neubauer, Matthew Feickert, Piero Altoe (NVIDIA), Kyle Cranmer, Maria Girone, Sofia Vallecorsa, Johannes Junggeburth, Nicolas Köhler, Jonas Graw, Michael Kagan, Mike Sokoloff, Daniele Bonacorsi, Gilles Louppe, Mario Campanelli, Michela Paganini, Paul Seyfert, Steven Schramm, Paolo Calafiura, Michele Floris, Jamal Rorie, Filip Siroky, Przemysław Karpiński, Alexander Radovic, Kim Albertsson, Jean-Roch Vlimant, Vladimir Vava Gligorov, Martin Erdmann, Stefano Carrazza, Zahari Kassabov

%\subsection{Introduction}
% Editors:
This chapter describes the science drivers and high-energy physics challenges where machine learning can play a significant role in advancing the current state of the art.
These challenges are selected because of their relevance and potential and also due to similarity with challenges faced outside the field.
Despite similarities, major R\&D work will go in adapting and evolving such methods to match the particular HEP requirements.

\subsection{Simulation}
\label{sec:fast-simulation}
% Editors: Michael Kagan, Matthew Feickert, Lukas Heinrich

Particle discovery relies on the ability to accurately compare the observed detector response data with expectations based on the hypotheses of the Standard Model or models of new physics.
While the processes of subatomic particle interactions with matter are known, it is intractable to compute the detector response analytically.
As a result, Monte Carlo simulation tools, such as GEANT~\cite{GEANT4}, have been developed to simulate the propagation of particles in detectors to compare with the data.
The dedicated CWP on detector simulation~\cite{simulationCWP} discusses the challenges of simulations in great detail.
This section focuses on the machine learning related aspects.\\

%Currently, the physics of proton-proton interactions are encoded in Monte Carlo generators, which produce a set particles that propagate through the detector material.
%However, the interactions of particles with the detector material can not be encoded in a single simple parametric model, and instead individual subatomic interactions are simulated with Monte Carlo methods.

For the HL-LHC, on the order of trillions of simulated collisions are needed in order to achieve the required statistical accuracy of the simulations to perform precision hypothesis testing. However, such simulations are highly computationally expensive. For example, simulating the detector response of  a single LHC proton-proton collision event takes on the order of several minutes. A particularly time consuming step is the simulation of particles incident on the dense material of a calorimeter. %, the detector used to measure the energy deposited by the particles. Radiative and nuclear interactions result in the production of a multitude of secondary particles, collectively referred to as a shower.
The high interaction probability and resulting high multiplicity in the so-called showers of particles passing through the detector material make the simulation of such processes very expensive. This problem is further compounded when particle showers overlap, as is frequently the case in the core of a jet of particles produced by high energy quarks and gluons.\\

Fast simulations replace the slowest components of the simulation chain with computationally efficient approximations. Often such approximations have been done by using simplified parametrizations or particle shower look-up tables. These are computationally fast but often suffer from insufficient accuracy for high precision physics measurements and searches.\\

Recent progress in high fidelity fast generative models, such as GANs and VAEs, which learn to sample from high dimensional feature distributions by minimizing an objective that measures the distance between the generated and actual distribution, offer a promising alternative for simulation.
%The application of such techniques to simulation of LHC data may provide a computationally efficient approach, which, unlike the simplified models used previously, can capture subtle correlations in the feature space, providing a reasonably-high precision simulation.
%
%The development of generative models for fast simulation may provide both a fast and high enough accuracy simulations for use at the HL-LHC.
A simplified first attempt at using such techniques saw orders of magnitude increase in simulation speed over existing fast simulation techniques~\cite{Paganini:2017hrr}, but such generative models have not yet reached the required accuracy partly due to inherent shortcomings of the methods and the instability in training of the GANs.  Developing these techniques for realistic detector models and understanding how to reach the required accuracy is still needed. The fast advancement in the ML community of such techniques makes this a highly promising avenue to pursue.\\

\medskip

% Although fast simulation is a necessity, some data analyses will require the highest fidelity simulations from GEANT.
Orthogonal to the reduction of demand of computing resources with fast simulations, machine learning can also contribute to other aspects of the simulation. Event generators have a large number of parameters that can be used to tune various aspects of the simulated events. Performing such tuning over many-dimensional parameter space is highly non-trivial and may require generating many data samples in the process to test parameter space points.  Modern machine learning optimization techniques, such as Bayesian Optimization, allow for global optimization of the generator without detailed knowledge of its internal details~\cite{Ilten:2016csi}. Applying such techniques to simulation tuning may further improve the output of the simulations.

%    \item Generative models (e.g. GANs), autoregressive models (Pixel RNN, etc)
%    \item Could imagine some use of ML in helping estimate the uncertainty envelope of the
%    simulations
%    \item Bayesian optimization \& sequential design for dramatically more efficient generation of monte carlo. As opposed to generating MC a priori, we can sometimes be much more efficient if we generate MC on the fly in areas where we need it.
%   Generative networks like CaloGAN, will help with that. Another promising avenue would be (as we think it is   written above) to replace I/O intensive pile-up to simulate the underlying event with ``just-in-time'' simulation of min bias background events (or the whole min-bias    background) using generative networks.
%    \item Improvement of MC integration techniques to vastly increase accuracy and scalability of multi-leg or very high order QCD processes. The reduction of the number of calls to a matrix element during integration by learning the shape of the matrix element in a multidimensional phase space with various poles considerably improves the speed of MC integration and corresponding unweighting. This allows for the generation of vast and theoretically accurate data samples which will be needed for the HL-LHC.

\subsection{Real Time Analysis and Triggering}
\label{sec:real-time-analysis}
% Editors: Vava + Conor

The traditional approach to data analysis in particle physics assumes that the interesting events recorded by a detector can be selected in real-time (a process known as \emph{triggering}) with a reasonable efficiency, and that once selected, these events can be affordably stored and distributed for further selection and analysis at a later point in time.
However, the enormous production cross-section and luminosity of the LHC mean that these assumptions break down.\footnote{They may well also break down in other areas of high-energy physics in due course.}
In particular there are whole classes of events, for example beauty and charm hadrons or low-mass dark matter signatures, which are so abundant that it is not affordable to store all of the events for later analysis. To exploit the full information the LHC delivers, it will increasingly be necessary to perform more of the data analysis in real-time~\cite{1742-6596-664-8-082004}.\\

This topic is discussed in some detail in the Reconstruction and Software Triggering chapter~\cite{recoCWP}, but it is also an important driver of machine learning applications in HEP. Machine learning methods offer the possibility to offset some of the cost of applying reconstruction algorithms, and may be the only hope of performing the real-time reconstruction that enables real-time analysis in the first place. For example, the CMS experiment uses boosted decision trees in the Level 1 trigger to approximate muon momenta. One of the challenges is the trade-off in algorithm complexity and performance under strict inference time constraints. In another example, called the HEP.TrkX project, deep neural networks are trained on large resource platforms and subsequently perform fast inference in online systems.\\

Real-time analysis poses specific challenges to machine learning algorithm design, in particular how to maintain insensitivity to detector performance which may vary over time. For example, the LHCb experiment uses neural networks for fast fake-track and clone rejection and already employs a fast boosted decision tree for a large part of the event selection in the trigger~\cite{Gligorov:2012qt}. It will be important that these approaches maintain performance for higher detector occupancy for the full range of tracks used in physics analyses. Another related application is speeding up the reconstruction of beauty, charm, and other lower mass hadrons, where traditional track combinatorics and vertexing techniques may become too computationally expensive.\\

In addition, the increasing event complexity particularly in the HL-LHC era will mean that machine learning techniques may also become more important to maintaining or improving the efficiency of traditional triggers. Examples of where ML approaches can be useful are the triggering of electroweak events with low-energy objects; improving jet calibration at a very early stage of reconstruction allowing jet triggers thresholds to be lowered; or supernovae and proton decay triggering at neutrino experiments.

%Many of the overabundant signals which must be analyzed in real-time are also the most computationally expensive to reconstruct in the first place. %This also connects to the increasingly heterogeneous computer architectures, since the most cost-effective architecture may be different for each algorithm, and the real-time analysis has to fit within a fixed budget so that the data are not lost.

%which requires rare quiet signals to of potentially noisy detectors.
%    \item di-Higgs (e.g. hh $\rightarrow$ 4b), for measuring self-coupling, triggering is a major problem
%    on low energy jets

\subsection{Object Reconstruction, Identification, and Calibration}
\label{sec:object-reco-id-calib}
% Editors: Michael Kagan, Matthew Feickert, Lukas Heinrich, Amir Farbin

The physical processes of interest in high energy physics experiments occur on time scales too short to be observed directly by particle detectors. For instance, a Higgs boson produced at the LHC will decay within approximately $10^{-22}$ seconds and thus decays essentially at the point of production. However, the decay products of the initial particle, which are observed in the detector, can be used to infer its properties. Better knowledge of the properties (e.g. type, energy, direction) of the decay products permits more accurate reconstruction of the initial physical process. Event reconstruction at large is discussed in~\cite{recoCWP} and also disucces the following applications of machine learning.\\

Experiments have trained ML algorithms on the features from combined reconstruction algorithms to perform particle identification for decades. In the past decade BDTs have been one of the most popular techniques in this domain. More recently, experiments have focused on extracting better performance with deep neural networks.\\

An active area of research is the application of DNNs to the output of feature extraction in order to perform particle identification and extracting particle properties~\cite{LHCbPID}.  This is particularly true for calorimeters or time projection chambers (TPCs), where the data can be represented as a 2D or 3D image and the problems can be cast as computer vision tasks, in which neural networks are used to reconstruct images from pixel intensities. These neural networks are adapted for particle physics applications by optimizing network architectures for complex, 3-dimensional detector geometries and training them on suitable signal and background samples derived from data control regions. Applications include identification and measurements of electrons and photons from electromagnetic showers, jet properties including substructure and b-tagging, taus and missing energy. Promising deep learning architectures for these tasks include convolutional, recurrent and adversarial neural networks. A particularly important application is to Liquid Argon TPCs (LArTPCs), which are the chosen detection technology for the flagship neutrino program.\\ %Algorithm reconstruction has proven to be difficult while early attempts at convolutional neural networks (CNNs) have shown better performance with little effort.

For tracking detectors, pattern recognition is the most computationally challenging step. In particular, it becomes computationally intractable for the HL-LHC. The hope is that machine learning will provide a solution that scales linearly with LHC collision density. A current effort called HEP.TrkX investigates deep learning algorithms such as long short-term memory (LSTM) networks for track pattern recognition on many-core processors.

\subsection{End-To-End Deep Learning}\label{subsec:endtoend}
\label{sec:applications-e2e}
The vast majority of analyses at the LHC use high-level features constructed from particle four-momenta, even when the analyses make use of machine learning. A high-profile example of such variables are the seven, so-called MELA variables, used in the analysis of the final states $\textrm{H} \rightarrow ZZ \rightarrow 4\ell$. While a few analyses, first at the Tevatron, and later at the LHC, have used the four-momenta directly, the latter are still high-level relative to the raw data. Approaches based on the four-momenta are closely related to the Matrix Element Method, which is described in the next section.\\

Given recent spectacular advances in image recognition based on the use of raw information, we are led to consider whether there is something to be gained by moving closer to using raw data in LHC analyses. This so-called \emph{end-to-end deep learning} approach uses low level data from a detector together with deep learning algorithms~\cite{Andrews:2018nwy,Andrews:2019faz}. One obvious challenge is that low level data, for example, detector hits, tend to be both high-dimensional and sparse.
Therefore, there is interest in also exploring automatic ways to compress raw data in a controlled way that does not necessarily rely on domain knowledge.

\subsection{Sustainable Matrix Element Method}
\label{sec:applications-MEM}
%Editor: M. Neubauer

% Need to add reference to the stand-alone whitepaper: HSF-CWP-018

The Matrix Element (ME) Method~\cite{Kondo:1988yd,Fiedler:2010sg,2011arXiv1101.2259V,Elahi:2017ppe} is a powerful technique which can be utilized for measurements of physical model parameters and direct searches for new phenomena.
It has been used extensively by collider experiments at the Tevatron for standard model (SM) measurements and Higgs boson searches~\cite{Abazov:2004cs,Abulencia:2006ry,Aaltonen:2008mv,Aaltonen:2010cm,Abazov:2009ii,Aaltonen:2009jj} and at the LHC for measurements in the Higgs and top quark sectors of the SM~\cite{Chatrchyan:2012xdj,Chatrchyan:2013mxa,Aad:2014eva,Khachatryan:2015tzo, Khachatryan:2015ila,Aad:2015gra,Aad:2015upn}. A few more details on the ME method are given in Appendix~\ref{subsec:MEM}.\\

%One can use calculations of Eq.~\ref{eqn:MEProb} in a number of ways to search for new phenomena at particle colliders. For measurement of model parameters $\boldsymbol\alpha$, one would maximize the likelihood function for observed events $\mathcal{L}(\boldsymbol\alpha)$ given by
%\begin{equation}
%\mathcal{L}(\boldsymbol\alpha) = \prod_{i} \sum_{k} f_k %\mathcal{P}_{\xi_k}({\bf x}_i|{\boldsymbol\alpha})
%\label{eqn:LH}
%\end{equation}
%where $f_k$ are the fractions of (non-interfering) processes contributing to the data. For new particle searches, one can (using Bayes' Theorem~\cite{Bayes01011763}) compute for a hypothesized signal $S$ the probability $P(S|{\bf x})$ given by
%\begin{equation}
%P(S|{\bf x}) = \frac{\sum_{i} \beta_{S_i} \mathcal{P}_{S_i}({\bf x}|\boldsymbol\alpha_{S_i}) }{\sum_{i} \beta_{S_i} \mathcal{P}({\bf x}|\boldsymbol\alpha_{S_i}) + \sum_{j} \beta_{B_j} \mathcal{P}({\bf x}|\boldsymbol\alpha_{B_j})}
%\label{eqn:LR}
%\end{equation}
%where, $S_i$ and $B_j$, denote all signal and background processes relevant to the considered phase space and $\beta$ are the \emph{a priori} expected process fractions. According to the Neyman-Pearson Lemma~\cite{Neyman289}, Eq.~\ref{eqn:LR} is the optimal discriminant function for $S$ in the presence of $B$ and can be used to extract a signal fraction in the data.

The ME method has several unique and desirable features, most notably it (1) does not require training data being an \emph{ab initio} calculation of event probabilities, (2) incorporates all available kinematic information of a hypothesized process, including all correlations, and (3) has a clear physical meaning in terms of the transition probabilities within the framework of quantum field theory.\\

One drawback to the ME Method is that it has traditionally relied on leading order (LO) matrix elements, although nothing limits the ME method to LO calculations. Techniques that accommodate initial-state QCD radiation within the LO ME framework using transverse boosting and dedicated transfer functions to integrate over the transverse momentum of initial-state partons have been developed~\cite{Alwall:2010cq}.
Another challenge is development of the transfer functions which rely on tediously hand-crafted fits to full simulated Monte-Carlo events.\\

The most serious difficulty in the ME method that has limited its applicability to searches for beyond-the-SM physics and precision measurements is that it is very \emph{computationally intensive}. If this limitation is overcome, it would enable more widespread use of ME methods for analysis of LHC data. This could be particularly important for extending the new physics reach of the HL-LHC which will be dominated by increases in integrated luminosity rather than center-of-mass collision energy.\\

The application of the ME method %Accurate evaluation of Eq.~\ref{eqn:MEProb}
is computationally challenging for two reasons: (1) it involves high-dimensional integration over a large number of events, signal and background hypotheses, and systematic variations and (2) it involves sharply-peaked integrands\footnote{a consequence of imposing energy/momentum conservation in the processes} over a large domain in phase space. Therefore,
despite the attractive features of the ME method and promise of further optimization and parallelization, the computational burden of the ME technique will continue to limit is range of applicability for practical data analysis without new and innovative approaches. The primary idea put forward in this section is to utilize modern \emph{machine learning techniques to dramatically speed up the numerical evaluations in the ME method} and therefore broaden the applicability of the ME method to the benefit of HL-LHC physics.\\

Applying neural networks to numerical integration problems is plausible but not new (see~\cite{CSEarticle2006,TICNC4344207,IJMC2013}, for example). The technical challenge is to design a network which is sufficiently rich to encode the complexity of the ME calculation for a given process over the phase space relevant to the signal process. Deep Neural Networks (DNNs) are strong candidates for networks with sufficient complexity to achieve good approximations, possibly in conjunction with smart phase-space mapping such as described in~\cite{Artoisenet:2010cn}. Promising demonstration of the power of Boosted Decision Trees~\cite{friedman2000,friedman2001} and Generative Adversarial Networks~\cite{GAN2014arXiv1406.2661G} for improved Monte Carlo integration can be found in~\cite{Bendavid:2017zhk}. Once a set of DNNs representing definite integrals is generated to good approximation, evaluation of the ME method calculations via the DNNs will be very fast. These DNNs can be thought of as preserving the essence of ME calculations in a way that allows for fast forward execution. They can enable the ME method to be both \emph{nimble} and \emph{sustainable}, neither of which is true today.\\

The overall strategy is to do the expensive full ME calculations as infrequently as possible, ideally once for DNN training and once more for a final pass before publication, with the DNNs utilized as a good approximation in between. A future analysis flow using the ME method with DNNs might look something like the following: One performs a large number of ME calculations using a traditional numerical integration technique like {\sf VEGAS}~\cite{PETERLEPAGE1978192,Ohl:1998jn} or {\sf FOAM}~\cite{JADACH200355} on a large CPU resource, ideally exploiting acceleration on many-core devices. The DNN training data is generated from the phase space sampling in performing the full integration in this initial pass, and DNNs are trained either \emph{in situ} or \emph{a posteriori}. The accuracy of the DNN-based ME calculation can be assessed through this procedure. As the analysis develops and progresses through selection and/or sample changes, systematic treatment, etc., the DNN-based ME calculations are used in place of the time-consuming, full ME calculations to make the analysis nimble and to preserve the ME calculations. Before a result using the ME method is published, a final pass using full ME calculation would likely be performed both to maximize the numerical precision or sensitivity of the results and to validate the analysis evolution via the DNN-based approximations.\\

There are several activities which are proposed to further develop the idea of a Sustainable Matrix Element Method. The first is to establish a cross-experiment group interested in developing the ideas presented in this section, along with a common software project for ME calculations in the spirit of~\cite{MoMEMta}. This area is very well-suited for impactful collaboration with computer scientists and those working in machine learning. Using a few test cases (e.g. $t\bar{t}$ or $t\bar{t}h$ production), evaluation of DNN choices and configurations, developing methods for DNN training from full ME calculations and direct comparisons of the integration accuracy between Monte Carlo and DNN-based calculations should be undertaken. More effort should also be placed in developing compelling applications of the ME method for HL-LHC physics. In the longer term, the possibility of Sustainable-Matrix-Element-Method-as-a-Service (SMEMaaS), where shared software and infrastructure could be used through a common API, is proposed.

\subsection{Matrix Element Machine Learning Method}
The matrix element method is based in the fact that the physics of particle collisions is encoded in the distribution of the particles' four-momenta and with their flavors. As noted in the previous section, the fundamental task is to approximate the left-hand side of Eq.~(\ref{eqn:MEProb}) for all (exclusive) final states of interest. In the matrix element method, one proceeds by approximating the right-hand side of Eq.~(\ref{eqn:MEProb}). But, since the goal is to compute $\mathcal{P}_{\xi}({\bf x}|{\boldsymbol\alpha})$, and given that billions of fully simulated events will be available, and that the simulations use exactly the same inputs as in the matrix element method, namely, the matrix elements, parton distribution functions, and transfer (or response) functions, one can ask whether a more direct machine learning approach can be developed to approximate  $\mathcal{P}_{\xi}({\bf x}|{\boldsymbol\alpha})$ without the need to execute the calculation of the right-hand side of Eq.~(\ref{eqn:MEProb}) explicitly. We believe the answer is yes, provided that a key advantage of the matrix element method can be replicated, namely, the fact that the method provides a function that depends explicitly on the model parameters ${\boldsymbol\alpha}$. Simulated events are typically simulated at fixed values of the model parameters. In order to replicate the advantage of the matrix element method, it would seem necessary to simulate events over an ensemble of model parameter points. But, that is a huge computational burden. The question is: is there a way to sidestep? Perhaps.\\

The matrix element method is technically feasible because there are now many codes that provide access to the square of the matrix elements as a function of ${\boldsymbol\alpha}$. Consequently, it would be possible to build a parametrized set of simulated events by reweighting each simulated event using the weighting function
\begin{align}
 w(\boldsymbol{\alpha}, \boldsymbol{\alpha}_0) & = \frac{|\mathcal{M}_{\xi}({\bf y}|\boldsymbol\alpha)|^2}{|\mathcal{M}_{\xi}({\bf y}|\boldsymbol\alpha_0)|^2},
\end{align}
where $\boldsymbol{\alpha}_0$ denotes the values of parameters at which the events were simulated. In practice, in order to keep the dynamic range of the weights within reasonable bounds, one presumably would simulate sets of events at a few reasonable choices of the parameters and use the weights to interpolate between these fixed parameter points.\\

What could one do with these parametrized simulated events? One could take advantage of the mathematical fact (shown more than quarter century ago) that, given a sufficiently expressive parametrized function $f(x, \omega)$, with parameters $\omega$, fitted by minimizing either the quadratic loss or cross entropy using data comprising two classes of objects of equal size, for example, signals with density $s(x)$ assigned a target of unity and backgrounds with density $b(x)$ assigned a target of zero, then asymptotically --- that is, for arbitrarily large training samples,
\begin{align}
 f(x, \omega) & = \frac{s(x)}{s(x) + b(x)}.
 \label{eqn:discriminant}
\end{align}
This result was derived in the context of neural networks. However, it is in fact entirely independent of the nature of the function
$f(x, \omega)$.\\

We can exploit this result to approximate $\mathcal{P}_{\xi}({\bf x}|{\boldsymbol\alpha})$ directly using, for example, DNNs. For each simulated event, in the training set, one could sample $\boldsymbol{\alpha}$ from a known distribution $q(\boldsymbol{\alpha})$, thereby yielding an ensemble of triplets $\{ ({\bf x}, \boldsymbol{\alpha}, w(\boldsymbol{\alpha}, \boldsymbol{\alpha}_0)) \}$. Call this ensemble the ``signal''.
Sample ${\bf x}$ from another known distribution $p({\bf x})$ and for each ${\bf x}$ sample $\boldsymbol{\alpha}$ from $q(\boldsymbol{\alpha})$. Call the ensemble of pairs $\{({\bf x}, \boldsymbol{\alpha})\}$ the ``background''.
From Eq.~(\ref{eqn:discriminant}), we have
\begin{align}
 f(x, \omega) & = \frac{s({\bf x}, \boldsymbol{\alpha})}{s({\bf x}, \boldsymbol{\alpha}) + p({\bf x}) q(\boldsymbol{\alpha})},
\end{align}
from which we find
\begin{align}
 \mathcal{P}_{\xi}({\bf x}|{\boldsymbol\alpha}) & = \frac{s({\bf x}, {\boldsymbol\alpha})}{q(\boldsymbol{\alpha})} = p({\bf x}) \left( \frac{f}{1 - f} \right).
\end{align}
If this could be made to work, it would be a direct machine learning alternative to the matrix element method, which incorporates the advantages of the former with the added advantage that the transfer function is automatically incorporated without the need to model it explicitly and the calculation of $\mathcal{P}_{\xi}({\bf x}|{\boldsymbol\alpha})$ would be fast because it would be given directly in terms of the DNN, $f(x, \omega)$.

\subsection{Learning the Standard Model}

New physics may manifest itself as unusual or rare events. One approach is to accurately identify the Standard Model processes and search for anomalies. Classifying the Standard Model events is a challenging task, as it consists of many complicated physics processes. Multi-class machine learning algorithms are well-suited for this classification problem. Once an event is classified as likely to be from a known physics process, it can be filtered out, and remaining events can be further analyzed for hints of new physics. Additionally, unsupervised machine learning techniques can be applied to remaining events to cluster them together. This approach would also be useful in identifying detector problems.
Another possibility is to use unsupervised learning to estimate the distributions of Standard Model events in the high-dimensional feature space. The comparison of such distributions with the measured datasets allows one to assess the compatibility of data with the background or to spot differences. The latter case would represent a hint of the presence of new physics events in the data. Density estimation techniques may be particularly useful in this task.

\subsection{Theory Applications}

The theoretical physics community has a number of challenges where machine learning can have an impact. These include areas of theoretical model optimization with hundreds of parameters, searches for new models,  understanding and estimation of the parton distribution functions and possibly quantum machine learning. The following details one such application: learning of the parton distribution functions with machine learning.\\

Making progress towards the objectives of the HL-LHC program (see Section~\ref{sec:introduction}) requires not only
obtaining experimental measurements of the physical processes but also reliable theory inputs to compare to. This becomes increasingly challenging as the experimental data gets more precise. There are numerous examples of phenomenologically relevant processes where the experimental uncertainty is comparable to the estimate of the theoretical uncertainty of the corresponding calculation.\\

Furthermore, the theory does not predict the value of all the inputs required for the computations (for example the
value of the strong coupling constant $\alpha_S$ evaluated at the $Z$ mass), and there are situations where the equations
resulting from theory cannot be solved to describe the physics adequately, and the corresponding theory inputs must be obtained from data instead. A more complex example is the determination of Parton Distribution Functions (PDFs): Quantum Chromodynamics (QCD) describes the proton collisions at high energy in terms of \emph{partons} (e.g. quarks and gluons), but it is not possible to calculate directly from QCD the momentum carried by each quark or gluon within a proton since QCD is not solvable in its confined regime. Our lack of theoretical knowledge about the characterization of
partons within a proton is embedded into a suitable definition of the Parton Distribution Functions (approximately the momentum densities of each of the partons). The PDFs then need to be determined from experimental data. The NNPDF collaboration uses machine learning techniques to obtain a PDF determination that is accurate enough to be suitable for high-precision collider data comparison. The NNPDF fitting procedure is described in full details in~\cite{Ball:2014uwa}.\\

The idea is to combine data from all relevant physical processes and fit a neural network representing each PDF. The difficulty of the procedure stems from the fact that multiple experimental inputs need to be combined to obtain
a PDF fit. Each of these inputs adjoins only indirect constraints on the PDFs, leaving some regions of the PDF
completely unconstrained by data. NNPDF fit includes around 50 datasets from different physical processes, and results that are not always consistent among themselves. Therefore, it is crucial to propagate the uncertainty of the experimental inputs into uncertainty on the PDFs via Monte Carlo Dropout ~\cite{2015arXiv150602142G}.\\

While the dataset is small, each experimental point has an indirect relation to the PDFs, as it is the result of the convolution of one or two PDFs with the corresponding partonic cross section. Code has been developed to compute these convolutions called \texttt{APFELgrid}~\cite{Bertone:2016lga}. Future research directions include the possibility of using standard ML frameworks to efficiently express the PDF fitting problem.
%The inclusion of high precision LHC data in the PDF fits presents new challenges in two different ways. First, the data
%can be precise enough that experimental and theory uncertainties are comparable in size. Consequently,
The uncertainties of the theory calculations need to be taken into account as well in the fits. A fully systematic treatment of theory errors in PDFs is a topic of research where machine learning could play an important role.  The dominant uncertainties in the data are no longer statistical and instead arise from correlated systematics. Determining those systematics accurately is non trivial on the side of the experimental analyses and can have a major impact on the
resulting PDFs. The problem grows more complex when ML techniques for which there is no simple recipe to estimate the uncertainty are used extensively in the experimental analysis. Taking full advantage of these advanced methods requires interdisciplinary research and communication on topics such as developing regularization schemes for experimental covariance matrices.\\

In conclusion, it is not only important to obtain the best fit PDF, but also a reliable estimation of the uncertainty, which in turn requires controlling the uncertainty of the experimental and theoretical inputs.

\subsection{Uncertainty Assignment}\label{sec:uncertainty}
Until fairly recently, little attention has been paid to the problem of assigning a measure of uncertainty to the output of machine learning algorithms. However, there is a growing recognition that the lack of such measures is a serious deficiency that needs to be addressed. This is particularly problematic in particle physics where accurate uncertainty assignments are crucial for assessing the quality of quantitative results. Uncertainty assignment is especially  urgent if machine learning methods are to be used for regression.\\

Standard methods exist in the statistics literature to quantify the uncertainty when an explicit likelihood function is available. Unfortunately, however, the overlap between the machine learning and statistics communities has been minimal at best.
The problem of uncertainty offers many opportunities for fruitful collaborations between physicists, statisticians, computer scientists, and machine learning practitioners, and is something that ought to be vigorously encouraged.
One area ripe for collaboration is in the use of Bayesian methods to quantify uncertainty using, for example, Hamiltonian Monte Carlo (HMC) sampling of posterior densities. One serious issue is that HMC is computationally prohibitive for DNNs. Therefore, new and more effective ways to perform large scale Bayesian calculations that go beyond HMC need to be developed, and it would be particularly useful to develop methods that can take advantage of multicore machines as discussed in Section~\ref{sec:resources}.
One possibility may be to explore deep probabilistic programming ~\cite{2017arXiv170103757T}

% and can be summarized as follows:

%\begin{itemize}
%\item Monte Carlo generation of artificial data.\\ Experimental data,
%  with central values, errors and their correlations are used to
%  generate further artificial data, consistent with the covariance
%  matrix provided by the experiment.
%\item Neural network fit to artificial data.\\ A genetic algorithm is
%  used to fit each artificial dataset to a neural
%  network representing the PDF.
%\item Predictions are later obtained by computing statistical
%  estimators (such as means, quantiles and standard deviations) over
%  the ensemble of neural networks.
%\end{itemize}

%A different direction of research concerns minimizing the computational expense of the PDF convolutions required to  (i.e. the cost to use the PDF sets). Several tools have been developed to produce reduced but statistically equivalent PDF sets~\cite{Carrazza:2016htc, Carrazza:2015aoa, Carrazza:2015hva}, and further steps are required to deliver the  PDFs efficiently encoded in binary form, instead of the text based LHAPDF format~\cite{Buckley:2014ana} currently in use.

\subsection{Monitoring of Detectors, Hardware Anomalies and Preemptive Maintenance}\label{sec:applications-monitoring}
Data-taking of current complex HEP detectors is continuously monitored by physicists taking shifts to check the quality of the incoming data. Typically, hundreds of histograms have been defined by experts and shifters are alerted when an unexpected deviation with respect to a reference occurs. It is a common occurrence for a new type of problem to remain unseen for a non-negligible period of time because such a situation had not been foreseen by the expert.\\

A whole class of ML algorithms called anomaly detection can be useful in such situations. They are able to learn from data and produce an alert when a deviation is seen. By monitoring many variables at the same, time such algorithms are sensitive to subtle signs forewarning of imminent failure, so that preemptive maintenance can be scheduled. Such techniques are already used in industry applications.\\

One challenge is that normal drifts in environmental conditions can induce drifts in the data. Beyond just reporting a problem, the natural next step is to connect an anomaly detection algorithm to the appropriate action: restart an online computer, contact an on-call expert, or similar. In the long term, the hardware and data structures of future detectors should be designed to facilitate the operation of anomaly detection algorithms.

\subsection{Computing Resource Optimization and Control of Networks and Production Workflows}\label{sec:resource-optimization}

% Editors: Daniele Bonacorsi, Valentin Kuznetsov
%\subsubsection{Application of ML for computing infrastructure}
%\label{sec:applications-infrastructure}
Data operations is one of the significant challenges for the upcoming HL-LHC. In the current infrastructure, LHC experiments rely on in-house solutions for managing the data. While these approaches work reasonably well today, machine learning can help automate and improve the overall system throughput and reduce operational costs.\\

Machine learning can be applied in many areas of computing infrastructure, workflow and data management. For example, dataset placement optimization and reduction of transfer latency can lead to a better usage of site resources and an increased throughput of analysis jobs. One of the current examples is predicting the ``popularity'' of a dataset from dataset usage, which helps reduce disk resource utilization and improve physics analysis time turn-over.\\

Data volume in data transfers is one of the challenges facing the current computing systems as thousand of users need to access thousands of datasets across the Grid. There is an enormous amount of metadata collected by application components, such as information about failures, file accesses, etc. Resource utilization optimization based on this data, including Grid components and software stack layers, can improve overall operations. Understanding the data transfer latencies and network congestion may improve operational costs of hardware resources.\\

Networks are going to play a crucial role in data exchange and data delivery to scientific applications in HL-LHC era. The network-aware application layer and configurations may significantly affect experiment's daily operations. ML can be applied to network security in order to identify anomalies in network traffic; predict network congestion; detect bugs via analysis of self-learning networks, and optimize WAN paths based on user access patterns.\\

\section{Collaborating with other communities}
\label{sec:collaboration}
% ---------------------------------------------------
% CHAPTER SUMMARY MESSAGE: < WRITE THIS BEFORE ANY EDITING>
% ---------------------------------------------------

%Contributors: Igor Lakomov (ALICE), Konstantin Kanishchev (AMS-02), Konstantin Skazytkin
%(ALICE), Jonas Eschle (LHCb), Alexander Kurepin (ALICE), Andrey Ustyuzhanin (Yandex,
%LHCb), Michela Paganini (ATLAS), Gabriel Perdue (MINERvA), Ryan Reece (ATLAS),
%Sean-Jiun Wang (CMS), Sergei Gleyzer (CMS), Meghan Kane (Industry - SoundCloud), Steven Schramm (ATLAS), Eduardo Rodrigues (LHCb), Martin Erdmann, Patrick Koppenburg (Nikhef, LHCb)

\subsection{Introduction}

Discovery science provides a challenge that attracts brilliant minds eager to push the boundaries of scientific understanding of nature. Particle physics has a rich problem domain that offers avenues for intellectual reward. The goal is to achieve a vibrant collaboration between the data science and high-energy physics communities by finding a common language and working together to further science.
%Before building collaborations, it is important to have a coherent strategy and vision of how to engage other communities.
Both communities can benefit from such collaboration. The HEP community can explore new research directions and applications of machine learning, novel algorithms, and direct collaboration on HEP challenges. The ML community can benefit from a diverse set particle physics problems with unique challenges in scale and complexity, and a large community of researchers that can expand the machine learning horizon by contributing to solving problems relevant to both communities. For example, the treatment of systematic uncertainties is an important topic for the HEP and ML communities. By working together on common challenges, the two fields can further progress in solving such problems.\\

There are a number of existing examples of collaboration between HEP and ML that have produced fruitful results through mostly local connections (e.g.~\cite{Likhomanenko:2016tgu,Paganini:2017hrr,darkmachines} ). The HEP community should continue such collaborations and look for additional collaborations with the ML community.\\

The HEP community needs to define its challenges in a language that the ML community can understand. This may involve only retaining necessary information with clear and concise explanations as to its relevance. Machine learning likewise has a significant amount of domain knowledge. Ideas and solutions provided by both communities should be presented in an understandable way for scientists without in-depth knowledge.

\subsection{Academic Outreach and Engagement}

Direct collaboration between HEP researchers and computer scientists working in the area of machine learning is an important possible driver of innovation in the area of machine learning applications described in Section~\ref{sec:applications}.
It is important for high-energy physics to engage and collaborate with the academic community focused on new machine learning algorithms and applications as they will naturally be interested in applying new ideas to interesting and complex data provided by HEP.\\

Conferences and workshops are a core aspect of the academic ML community, and organizing or contributing to key conferences is a means of gaining interest.
Organizing sessions or mini-workshops within major ML conferences, such as NIPS, would increase the familiarity of HEP within the ML community and jump-start future collaborations.
This has been explored in single cases~\cite{NIPS:2015:ALEPH} but is not an established, regular workshop series.
At the same time inviting ML experts to HEP workshops, as done at~\cite{FlavourDataMining} and the DS@HEP series~\cite{DSatHEP2015, DSatHEP2016, DSatHEP2017}, can foster greater long-term collaboration.
There should be coordinated efforts to:
\begin{itemize}
 \item Organize workshops and conferences open to external collaborators to discuss applications, algorithms and tools
 \item Organize thematic workshops around topics relevant to HEP
       %\item Engage machine learning experts by creating prestigious fellowships and data-science related initiatives
\end{itemize}

%\subsection{Science outreach}
HEP should reach out to other scientific communities with similar challenges, for example astrophysics/cosmology, medium energy nuclear physics and computational biology. This can lead to more active partnerships to better collaborate on ideas, techniques, and algorithms.

%[ A comment : I see that wrt the version I printed (June) the scope has been reduced to reflect what is actually described in Sec.6: collaboration between HEP and ML communities. I find that a bit sad as one could (should?) think of other fields of science that use ML a lot. Astrophysics is mentioned above, but social sciences and genomics face similar challenges. A quick search for ML in bioRxiv is enlightening. Could this section be expanded a bit?  ]

\subsection{Machine Learning Challenges}

To engage the wider ML community, challenges such as the Higgs Boson Challenge (2014) or the Flavor Physics Challenge~\cite{NIPS:2015:ALEPH,FlavourDataMining} have been organized on Kaggle.
These types of challenges draw considerable attention from the machine learning community and additional similar challenges should be organized in the future.\\

Organization of a challenge requires a well documented dataset, a starting-kit and an evaluation metric to rank the solutions.
This forces the organizers to simplify the problem as much as possible, while retaining its intrinsic complexity.
The drawback of challenges is that once they are launched, participants priority is winning the challenge and not eventual collaboration with HEP.
It is important to foresee upfront a way to integrate incoming solutions, for example via forums and post-challenge workshops (like~\cite{FlavourDataMining}) where a diversity of competitive algorithms can be presented.
The challenge dataset and evaluation metric should be released publicly so that further developments can continue.

\subsection{Collaborative Benchmark Datasets}\label{subsec:benchmark}
There is a strong incentive for HEP to develop public benchmark datasets, beyond just challenges. Access to a dataset makes the discussion much more concrete and productive. Within the HEP community, common datasets enable comparisons of algorithms with much better accuracy, which is very useful for research and development. The same benchmark datasets can also be used for teaching, tutorials and training.\\

These benchmark datasets could be built based on public simulation engines or released by experiments within the bounds of their data access policy. Even a small subset of an experiment's simulated data can be the base of a very valuable benchmark dataset. For example, the CMS experiment has released a significant amount of its simulated and collected data via the CERN Open Data Portal~\cite{opendata}.\\

To be maximally useful, the subsequent guidelines should be followed when designing a benchmark dataset:
\begin{itemize}
 \item Simplify the dataset as much as possible while conserving all methodical difficulties relevant to solving the problem
 \item Document the dataset to make it understandable by a non-HEP expert
 \item Create methodologies and metrics for the evaluation of proposed solutions, and document them
 \item Prepare an integration plan for incoming ideas and solutions
 \item Feedback results of successful applications
\end{itemize}

%\subsection{Collaborative Benchmark Datasets}

%As discussed in Section~\ref{subsec:benchmark}, collaborative benchmark datasets can be useful for developing ML challenges and for collaboration with the ML community.

The HEP community should organize and curate a variety of such benchmark datasets covering its current physics drivers and make them publicly available. To improve the reproducibility of results and algorithm comparisons, some of the data used for evaluation of the solutions should be kept private.
%an effort known as TrackingML is underway to create datasets and establish a challenge to elicit solutions from the broader ML community, Move to bridges.
Additionally, after investing heavily into producing highly-detailed and realistic simulations, the HEP community can provide the machine learning community with labeled datasets with high statistical power to test algorithms and develop novel ideas.

%\subsection{Communication and Outreach}
%A multi-prong strategy for engaging the ML community is presented, targeting different areas: academia, industry and the ``general public''.

%\subsection{Inter-experimental Collaboration}
%to be added

\subsection{Industry Engagement}
Industry has been focused on the development and adoption of machine learning techniques. In addition to algorithm and software development, one of the promising areas is the adoption of dedicated specialized hardware and high performance co-processors. GPUs, FPGAs, and high core count co-processors all have the potential to dramatically increase performance of machine learning applications relevant to the HEP community.
One of the challenges is gaining the human expertise for development and implementation. Industry interactions bring specific technology opportunities and access to specialized expertise that can be difficult to hire and support internally.\\

There are specific areas of development where industry has expressed interest in collaborating with HEP.  Automated resource provisioning, data placement, and scheduling are similar to industrial applications to improve efficiency. Applications such as data quality monitoring, detector health monitoring and preventative maintenance can be automated using techniques developed for other industrial quality control applications. There are two more forward looking areas that coincide with HEP physics drivers, namely computer vision techniques for object identification and real-time event classification. These present a challenge to industry due to its complexity and benefit outside of HEP.\\

%		\item Bridge to industry (partnerships in applying externally developed techniques and tools to HEP problems anomaly detection against fraud transactions, spam filtering etc)

%\subsubsection{CERN OpenLab and research-industry collaborative initiatives}
CERN OpenLab is a public-private partnership that accelerates the development of cutting-edge solutions for the LHC community and wider scientific research. CERN OpenLab has established the infrastructure to maintain non-disclosure agreements, to arrange ownership of intellectual property, and to provide an interface between CERN and industry. As part of its upcoming phases, OpenLab plans to explore machine learning applications for the benefit of LHC experiments computing and the HL-LHC. Such initiatives and industry partnerships should be supported in the future.

\subsection{Machine Learning Community-at-large Outreach}
Another form of engagement is using the communications mediums to broadcast our challenges and attract interested collaborators. There are a variety of channels which can be leveraged to increase the visibility of our problems and research opportunities in the ML community. These can be popular forums such as reddit, personal or official blogs, social media, or direct contact with influential personalities.\\

%\subsubsection{Podcasts, Blog Posts and Meetups}
Podcasts have shown to be a great vehicle for reaching a large audience. Listeners are keen to consume material that is outside of their immediate problem domain in a way that is easy to digest. There is an abundance of machine learning podcasts with a large base of listeners that can be targeted for outreach:
\begin{itemize}
 \item \href{http://lineardigressions.com/}{\textcolor{blue}{Linear Digressions}} (co-hosted by former ATLAS Ph.D. Katie Malone)
 \item \href{http://partiallyderivative.com/}{\textcolor{blue}{Partially Derivative}}
 \item \href{http://www.thetalkingmachines.com/}{\textcolor{blue}{Talking Machines}}
 \item \href{https://dataskeptic.com/}{\textcolor{blue}{Data Skeptic}}
 \item \href{http://becomingadatascientist.com/}{\textcolor{blue}{Becoming a Data Scientist Podcast}}
 \item \href{https://itunes.apple.com/us/podcast/not-so-standard-deviations/id1040614570?mt=2}{\textcolor{blue}{Not So Standard Deviations}}
 \item \href{https://twimlai.com/}{\textcolor{blue}{This Week in ML \& AI}}
\end{itemize}

%Some examples are (Partially Derivative Episode: \href{http://partiallyderivative.com/podcast/2017/02/28/michael-kagan}{``Particle Physics and Machine Learning at CERN with Michael Kagan''}) as well as applications of programming and technology in physics (Talk Python To Me Episode \#29: ``\href{https://talkpython.fm/episodes/show/29/python-at-the-large-hadron-collider-and-cern}{Python at the Large Hadron Collider and CERN}'' with Kyle Cranmer).

%Bringing together tangentially related fields is commonplace in ML podcasts. For example, podcasts such as This Week in Machine Learning (TWiML) features machine learning experts across a wide variety of problem domains, spanning topics from a  \href{https://twimlai.com/twiml-talk-12-brendan-frey-reprogramming-human-genome-w-ai/}{\textcolor{blue}{genomics}} to \href{https://twimlai.com/twiml-talk-027-intelligent-autonomous-robots-ilia-baranov/}{\textcolor{blue}{robotics}} to \href{https://twimlai.com/twiml-talk-5-joshua-bloom-machine-learning-stars-productizing-ai/}{\textcolor{blue}{astrophysics}}.Podcasts typically consist of a casual conversation from 30 minutes to an hour. Therefore, there is a high yield in the form of listeners versus time investment required to release a podcast.
%\subsubsection{Blog Posts}

Another form of engagement is through outreach-style blog posts to explain HEP challenges in a way that is easy to understand by the public.\\

Another outreach opportunity is to make HEP related presentations at machine learning Meetups across the world to generate awareness, engage community, foster cross pollination of ideas between HEP and industry. Some popular ML meetups are:
\begin{itemize}
 \item NYC: \url{https://www.meetup.com/NYC-Machine-Learning/}
 \item Berlin: \url{https://www.meetup.com/Advanced-Machine-Learning-Study-Group/}
 \item SF: \url{https://www.meetup.com/SF-Bayarea-Machine-Learning/}
\end{itemize}

In conclusion, existing outreach efforts should be expanded to attract greater collaboration between the HEP and ML communities. By understanding and speaking the same language, the two communities can better collaborate and find solutions to present and future challenges.

%Section outline:
%\begin{itemize}
%	\item Describe the need for a bridge and its benefits
%	\item Explain community differences
%	\item ML rapid experimentation
%	\item ML and HEP Domain knowledge differs (need to be able to communicate)
%	\item Roadmap to creating a bridge
%	\item Define the problem
%	\item Discuss a common approach within HEP for reaching out to ML
%	\item Reach out to ML
%	\item Become more involved members of the ML community
%	\item Support new collaborations between HEP-ML
%	\item Make benefits of collaborations accessible to the community
%\end{itemize}

\section{Machine Learning Software and Tools}
\label{sec:software}
% ---------------------------------------------------
% CHAPTER SUMMARY MESSAGE: < WRITE THIS BEFORE ANY EDITING>
% ---------------------------------------------------

Machine learning does not exist without software. There are a large variety of algorithms written in different programming languages and general software frameworks that combine many classes of methods into one package. The following sections focus on specific topics and challenges related to machine learning software design in HEP.

%\subsection{HEP Machine-Learning Software}
%The HEP community has developed a baseline with which any new machine-learning tools or methods can be compared with.

%We are able to keep up and in some cases improve over external tools on HEP benchmarks. Internally-developed tools have GPU-capable deep-learning libraries. On the other hand, not all external data-formats, options, model architectures and variants are currently fully supported.

\subsection{Software Methodology}
Presently, there are two machine learning software methodologies in high-energy physics. The first approach is to implement abstract ML algorithms in HEP-developed toolkits, such as the Toolkit for Multivariate Analysis (TMVA) in ROOT. This provides on site support to physicists and dedicated development for HEP data formats and applications. The second approach is to rely on externally developed software, of which there are many examples. This often requires tedious and repetitive work to adapt HEP data formats to external software, breaks analysis workflows and introduces difficulties in the analysis software development. Historically, a variety of approaches and competition among them has led to important breakthroughs in the field. On the other hand, having too many choices increases repetition and leads community segmentation and possible issues with reproducibility.
There is some convergence of the two approaches. Converters have been written for the reintegration of some externally trained models into HEP tools, like~\cite{lwtnn}. Complementarily, interfaces between HEP and external tools have also been developed.%  More effort is needed to extend these interconnecting efforts and integrate new useful tools when they become available.

\subsection{I/O and Programming Languages}\label{sec:software_IO}
The sheer amount of data accumulated by HEP experiments requires data access optimization because. Such I/O performance is very dependent on data formats. Moreover, support for reading data in different formats is required for certain use-cases.\\
%In particular, data streaming, locality and partitioning will impact training and cost of data processing.

Exploration of new file systems and methods to improve I/O limitations are important and the following R\&D studies should take place:
\begin{itemize}
 \item Explore new file systems to assess I/O limitations;
 \item Use alternative industry approaches such as Google BigQuery to explore various data access patterns;
 \item Explore parallel data processing platforms such as Apache Spark for ML training.
\end{itemize}

Although particle physics has been reliant on C++ over the past decade, the machine learning community has explored other programming languages, in particular the python-based ecosystem.

\subsection{Software Interfaces to Acceleration Hardware}\label{sec:software_proglang}
Modern machine learning software significantly benefits from the use of hardware accelerators such as GPUs. At the same time, ML users should not be hindered in their development of new applications by writing platform-dependent code. Various interfaces to different hardware architectures are needed in order to make efficient use of the available computing resources. The emergence of the Open Computing Language (OpenCL) allows programming of high-level interfaces that can run on various hardware platforms.\\

Machine learning tools often provide different sets of APIs to develop and train the models in one language, and various bindings to use trained models in other programming languages. This is a convenient model for many HEP applications, such as the trigger system, where (1) application latency puts stringent requirements on the software and hardware used and (2) the general purpose training platform might be different from the highly specific deployment platform.

\subsection{Parallelization and Interactivity}
Training ML algorithms takes a significant amount of time and parallelization at various levels is desired, such as the parallelization of the computations within a single model. Another type of parallelism is data parallelism that targets the processing phase of the training with data partitioning and model training using distributed workers. Frameworks like Apache Spark and ideas such as batch training offer promise in this area.\\

One often needs to produce many different machine learning models, for example while tuning hyper-parameters or performing k-fold cross-validation, and distribution of these algorithms is key to the reduction of the overall training time.
ML algorithm inference significantly benefits from parallelization as well. Although event selection in a particle physics trigger system is ``embarassingly parallel~\cite{embarrassingly}'', parallel processing frameworks are developed to reduce the memory footprint. Complex models might well be too big to reside once per CPU core in memory and need to be hosted in shared memory. At the same time the stringent latency requirements impose constraints on the type of algorithms that can be easily deployed.\\

The availability of interactive frameworks, for example Jupyter notebooks, allows for rapid prototype development and testing of ML tools. Such frameworks also ease the connection between the description of models and the data, providing straightforward means of visualizing models and data.
HEP has started to exploring interactive frameworks, such as the Service for Web Based Analysis (SWAN)~\cite{swan}. One of the challenges is availability of adequate hardware resources for these systems.

%\subsection{Brief History of HEP Machine-Learning Software}
%\label{sec:software_history}

%\begin{description}
%\item[1991 - 1998]
%	\begin{itemize}
%		\item C. Peterson and T. Rognvaldsson, An Introduction to Artificial Neural Networks, Lectures given at the 1991 CERN School of Computing, University of Lund preprint LU TP 91-23, September 1991.
%		\item Early Feed-forward NN’s B. Denby, Neural Computation 5 (1993) 505.
%		\item Kernel Density http://inspirehep.net/record/407093
%		\item Fisher discriminants \& H-matrix http://inspirehep.net/record/1234704
%	\end{itemize}
%	\item[1998-2005] Initial HEP Machine Learning Software development based on external algorithms lead to number of standalone packages (MultilayerPerceptron and others)
%	\item[2005-2010] Abundance of comparably-performing methods lead to a consolidation into software toolkits such as Toolkit for Multivariate Analysis (TMVA) and StatPatternRecognition (SPR).
%	\item[2008] SPR stops development and support, TMVA takes over
%	\item[2010-2014] Wide use of TMVA in HEP, full integration in ROOT and switch to maintenance mode
%	\item[2014-present] Upgrade/modernization of TMVA, wider use of externally developed tools deep learning revolution.
%\end{description}

\subsection{Internal and External ML tools}
%Contributors: Juan Pedro Araque, Nuno Filipe Castro, Thomas Keck, Stefan Wunsch, Igor Lakomov, Konstantin Skazytkin, Luke Kreczko, Przemysław Karpiński, Claire David, Hans Pabst, Attilio Picazio, Giles Strong, Gordon Watts, Lorenzo Moneta, Matthew Feickert, Mark Neubauer, Daniele Bonacorsi, Mauro Verzetti, Seth Moortgat, Sergei Gleyzer, Fernanda Psihas, Elias Coniavitis, Gilles Louppe, Kim Albertsson, Steven Schramm, Martin Erdmann

% In Section~\ref{sec:data_formats} we provide a high-level overview of existing HEP and ML data formats. Section~\ref{sec:ml_tools} investigates the present ML software landscape in HEP, while Section~\ref{sec:tool_roadmap} proposes a road-map towards a sustainable future HEP machine learning software landscape.

Internally developed ML software, such as the Toolkit for Multivariate Analysis (TMVA,~\cite{TMVA}), has been developed to apply a variety of machine learning algorithms to HEP challenges. Currently, many published HEP analyses with machine learning have made use of TMVA. There is also software developed in HEP, such as NeuroBayes~\cite{neurobayes,neurobayes2} and RuleFit, that have gained popularity outside of HEP.\\

At the same time, the ML landscape has evolved and many different ML tools have emerged and gained popularity. There are a growing number of published results based on externally developed tools. The latter, often developed directly by industry for specific applications, are constantly undergoing development, incorporating the latest algorithms from academia. Currently, both internal and external tools are used by the HEP community. TMVA has also undergone significant development in recent years.
In addition, there are smaller tools developed in the HEP community, extending either internal or external ML tools for specific use cases and applications within HEP experiments, such as \texttt{hep\_ml}~\cite{hep_ml} for training ML classifiers in HEP specific tasks, or \texttt{tmva-branch-adder}~\cite{tmva-branch-adder} and \texttt{lwtnn}~\cite{lwtnn} to facilitate classifier inference tasks.\\

This begs the question: what aspects of ML development and use should the HEP community focus on in the next 5-10 years? There are several aspects to consider including data formats, community size, and interfaces.
%and middleware solutions aid in the conversion as needed.

\subsubsection{Machine Learning Data Formats}\label{sec:data_formats}

\begin{figure}
 \centering
 \includegraphics{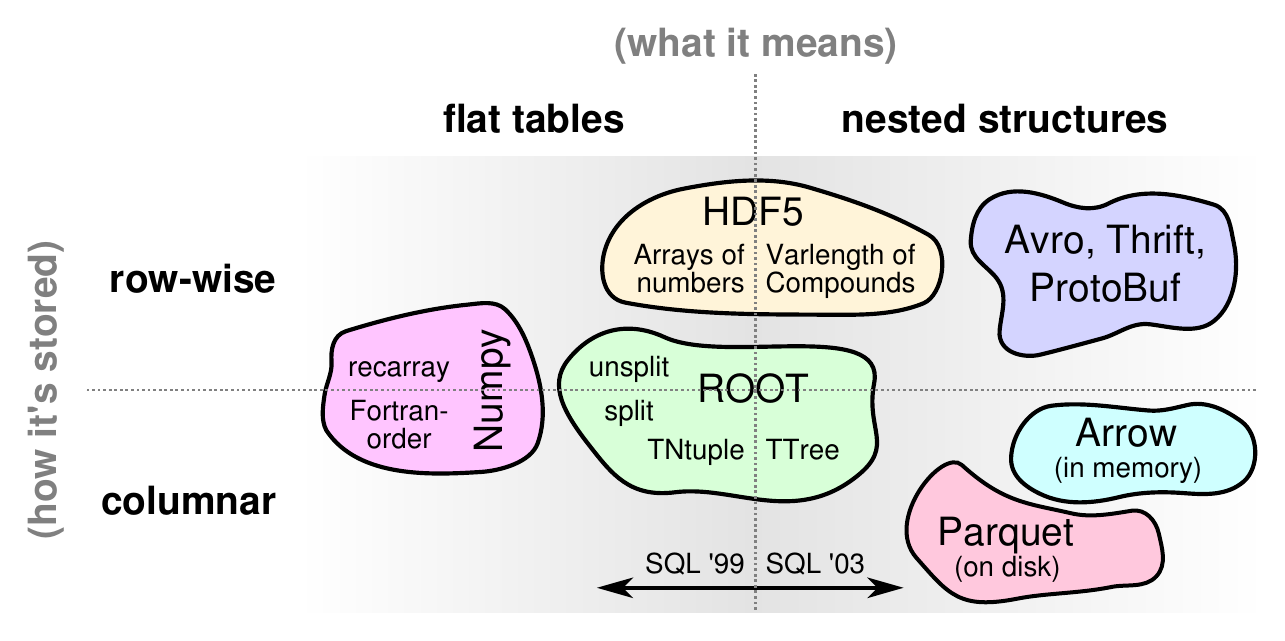}
 \caption{Existing data-formats used by ML communities.}\label{fig:data-formats}
\end{figure}

HEP and ML communities currently make use of different data formats. HEP heavily relies on the ROOT software framework for data storage, data processing, and data analysis. The machine learning community uses a large variety of formats, as shown in Figure~\ref{fig:data-formats}. This figure also shows the relationship of machine learning data formats with ROOT: the ROOT file format is very flexible, though it requires a significant time investment to learn the proper use. Table~\ref{tab:formats_vs_tools} summarizes the current machine learning toolkits and file formats they support.
% the use of the following data formats: flat tables, sparce matrices, row and column-wise arrays and static data structures, in various machine learning toolkits

\begin{table}
  \caption{This table lists various data formats (rows) and ML tools (columns). The \checkmark\ indicates that there is a native solution, while $\times$ means that the conversion from one data-format to another is straightforward (the conversion mechanisms from Table~\ref{table:Middleware} are omitted here). The following notation has been used to denote the data-formats: \textbf{T} Trees, \textbf{F} flat tables, \textbf{M} sparse matrices, \textbf{R} row-wise arrays, \textbf{C} column-wise arrays \textbf{S} static data structures\newline}
 %
 % Please don't comment in text, use hypothesis instead
 %\textcolor{red}{don't root\_numpy and root\_pandas provide straightforward conversion from root to numpy/hdf5 and thereby input formats to tensorflow, theano, scikit learn, sparkml, torch (which would fill the first row with $\times$)?}
 %
 %\centering
 \begin{tabular}{lcccccccccc}
  \hline
                          & TMVA              & TensorFlow & Theano     & Scikit     & R          & Spark      & VW         & libFM      & RGF        & Torch      \\
                          &                   &            &            & Learn      &            & ML         &            &            &            &            \\
  \hline
  \hline
  ROOT [\textbf{T, C}]    & \checkmark        & \multicolumn{9}{|c|}{through conversion into other formats, see Table~\ref{table:Middleware}}                      \\
   \cline{3-11}
  CSV [\textbf{F}]        &                   & \checkmark & \checkmark & \checkmark & \checkmark & \checkmark & $\times$   & $\times$   & $\times$   & \checkmark \\
  libSVM [\textbf{M}]     &                   &            &            &            &            &            & $\times$   & \checkmark & $\times$   &            \\
  VW [\textbf{M}]         &                   &            &            &            &            &            & \checkmark &            &            &            \\
  RGF [\textbf{M}]        &                   &            &            &            &            &            &            &            & \checkmark &            \\
  NumPy [\textbf{R}]      & \cite{root_numpy} & \checkmark & \checkmark & \checkmark & \checkmark & \checkmark & $\times$   & $\times$   & $\times$   & \checkmark \\
  Avro [\textbf{S, R}]    &                   &            &            &            & \checkmark & \checkmark &            &            &            &            \\
  Parquet [\textbf{S, C}] &                   &            &            &            & \checkmark & \checkmark &            &            &            &            \\
  HDF5 [\textbf{S}]       &                   & $\times$   & $\times$   & $\times$   &            &            &            &            &            & \checkmark \\
  R df [\textbf{R}]       &                   &            &            &            & \checkmark &            &            &            &            &            \\
  \hline
 \end{tabular}\label{tab:formats_vs_tools}
\end{table}
%The de-facto HEP data format is ROOT, used by current HEP experiments and software systems. Although it is very well suited to store and process physics events, it is seldom  used outside of HEP.
\subsubsection{Desirable HEP-ML Software and Data Format Attributes}

A desirable data format should have the following attributes: high read speed for efficient training, sparse readability without loading the entire dataset into RAM, compression, and common use by the machine learning community.\\

HEP machine learning applications require highly performant and flexible algorithms to address the variety of use cases. Some applications, such as triggering, also have to work under tight latency constraints of the order of a few microseconds and below. The data sets are extremely large, which comes with I/O challenges as described in Section~\ref{sec:software_IO}. This is expected to become even more challenging, as the LHC continues to ramp-up and deliver increasingly large amounts of data.\\

%All of these constitute requirements on ML tools used by the HEP community.
%There are further requirements. The HEP community, for example, is looking to elevate Python to a first class library.
As discussed in Section~\ref{sec:software_proglang}, machine learning tools use a number of languages. To use these tools it will therefore be important to offer adequate support. In production software environments, ML classifiers will be part of larger C++ or python applications. Classifiers from external tools will probably fit perfectly into python applications, but for C++ applications, restrictions to toolkits with C++ API or the use of C++ converters may be needed to make sure the training result can be evaluated.\\

Advantages of using the {\bf external tools} are the size of the community that uses and supports them, being able to easily keep up with progress in the industry and profit from the forefront of the ML research.
It should also be noted that some of the recent industrial efforts to develop and maintain ML-tools rely on resources far beyond that of basic research. The deep learning tools of the previous and current generations constitute a demonstration of corresponding quality.\\

Disadvantages of using external tools are that, they is often no guaranteed support over the lifetime of particle physics experiments, and it can be difficult to adapt them to HEP specific requirements which may not be among the priorities of the ML community.\\

Advantages of using {\bf internal tools} are that decisions about long-term support remain in the community, and the tools can be adapted to the specific needs of HEP. Disadvantages include that any new algorithm and idea first needs to be ported before their use can be evaluated.

\subsubsection{Interfaces and Middleware}

One approach to bridge the gap between internal and external tools is by building interfaces. Some researchers prefer to convert their data to the file formats used by external tools and to work exclusively with external tools. This has the advantage of working as close as possible with the ML community's tools and its documentation, and is as close to what is used by ML researchers as possible. This may come with the difficulty of converting back to a HEP format, when analyzing the data further with a fitting framework~\cite{RooFit, RooStats, HistFactory}.
At the same time, interfaces have been built between TMVA and external machine learning tools, allowing for their use and direct comparison between their performance. Currently, interfaces to R, scikit-learn, keras and tensorflow have been developed. Those have the advantage of providing a homogeneous interface and require little training overhead for those already knowledgable in using TMVA.\\

A more general approach to file format conversion is to build middleware solutions that export HEP-specific formats like ROOT to formats used by external machine learning tools.
%\textit{Caution: this middleware doesn’t imply that it hides completely the other layers. Here one needs knowledge on both inputs and outputs to make a bridge between the HEP tools and external tools.}
%As part of the effort to increase collaboration and bring the HEP community closer to the machine learning community, we foresee that providing a suitable middle-ware solutions to convert ROOT based data to various data-formats is a crucial component to collaboration.
Existing middleware solutions are shown in Table~\ref{table:Middleware}.\\

Approaches to bridge the different languages and data formats inside and outside of HEP include providing interfaces or building middleware solutions that translate HEP-specific data formats to external ML tools. It is a topic of current research to determine the most efficient solution.

\begin{table}
 \caption{Middleware solutions translating the ROOT data format to other formats.}
 \begin{center}
  \begin{tabular}{|l|l|}
   \hline
   \textbf{PyROOT}       & Python extension module that allows the user to interact with ROOT data/classes.~\cite{PyROOT}          \\
   \hline
   \textbf{root\_numpy}  & The interface between ROOT and NumPy supported by the Scikit-HEP community.~\cite{root_numpy}           \\
   \hline
   \textbf{root\_pandas} & The interface between ROOT and Pandas dataframes supported by the DIANA/HEP project.~\cite{root_pandas} \\
   \hline
   \textbf{uproot}       & A high throughput I/O interface between ROOT and NumPy.~\cite{uproot}                                   \\
   \hline
   \textbf{c2numpy}      & Pure C-based code to convert ROOT data into Numpy arrays                                                \\
                         & which can be used in C/C++ frameworks.~\cite{c2numpy}                                                   \\
   \hline

   \textbf{root4j}       & The hep.io.root package contains a simple Java interface for reading ROOT files.                        \\
                         & This tool has been developed based on freehep-rootio.~\cite{root4j}                                     \\
   \hline

   \textbf{root2npy}     & The \texttt{go-hep} package contains a reading ROOT files.                                              \\
                         & This tool has been developed based on freehep-rootio.~\cite{root4j}                                     \\
   \hline

   \textbf{root2hdf5}    & Converts ROOT files containing TTrees into HDF5 files containing HDF5 tables.~\cite{root2hdf5}          \\
   \hline
  \end{tabular}
 \end{center}
 \label{table:Middleware}
\end{table}

\section{Computing and Hardware Resources}
\label{sec:resources}
% ---------------------------------------------------
% CHAPTER SUMMARY MESSAGE: < WRITE THIS BEFORE ANY EDITING>
% ---------------------------------------------------
\vspace{10pt} % Make Chapter start after Table 2
A typical high-energy physics data model consists of a hierarchy of increasingly refined data stores. Each store provides a refined view of a list of ``events'', the self contained records that capture the state of the detector at the time when a particle interaction occurs. At the bottom of the hierarchy is the raw data, a byte-stream of the readout from detector electronics. At the top of the hierarchy are the ``high-level'' physics objects, such as electrons or jets, providing descriptive information about the quality and topology of physics events. The data stores are typically processed by independent copies of identical code processed in batch computing queues. The result of this processing is filtered data and extracted physics parameters.\\% like production cross sections, coupling constants, and background rates.

At present, training of machine learning algorithms is done using dedicated or private resources. These vary in configuration and processing power, depending on the size of the data and complexity of the algorithm. For a given event, the evaluation of algorithms is performed on a single core producing a single discriminator or regressor output. In order to progress to evaluation of complex machine learning algorithms, more computing power is needed in both the training and evaluation stages, as larger amounts of data are needed to feed models with tens or hundreds of thousands of parameters. This implies the expansion of the current computing model to include architectures that are well suited to machine learning tasks, such as many integrated core (MIC), graphical processing units (GPUs) and Tensor Processing Units (TPUs). This is a fundamental departure from the single-core or few-core jobs. These architectures provide a significant computational speed improvement for both training and evaluation of ML algorithms, but require dedicated hardware, drivers, and software configuration.\\

Similarly, the locality and bandwidth of large data stores will need to be optimized in order to avoid bottlenecks in training and evaluation for analysis. Data placement and the need to use dedicated hardware indicate that a transition to HPC, or HPC-like, architectures may be needed to achieve the desired performance. Due to significant synergy with the direction of industry in this respect, use of commercially available resources should be considered for future high-energy physics computing models.\\

In the following subsections we will discuss the resource needs for the physics drivers mentioned earlier: fast simulation, real-time analysis, object and event reconstruction and particle identification. The limitations of the current computing model are discussed as well as how those physics driver needs can be met in the future.

%In this section we will discuss the currently available resources for ML applications in the HEP community, and the necessary steps to take towards taking full advantage of ML within computing for high energy physics.

\subsection{Resource Requirements}

The popularity of deep learning methods is to a large extent due to the possibility of training these models in a reasonable amount of time with large scale parallelism. In particular, the training stage requires repeated simultaneous access to many data elements and specialized hardware has been developed for training deep learning models.\\ %Moreover, in future experiments, the datasets used to train ML algorithms will be of order petabytes or larger, requiring optimized access to those data to feed the algorithms.

In contrast, inference can be an operation applied to a single data element at a time and needs to be performed only once. Inference has less demands on I/O and is limited only by the computing power and model complexity. Because inference has real-time applications in high-energy physics, latency and throughput constraints are the main challenges.\\

A typical HEP application can require up to 1 GPU-week to train a single model.
To obtain the best results and understand the performance of the model, an average of 100 hyper-parameter points optimization is typically performed. A single project could therefore easily require up to a full GPU-year for training.
The speed up of the training process can be obtained by means of faster and more capable hardware, parallelization of single training and by splitting training over multiple nodes. Resources, such as GPUs, TPUs and MIC need to be evaluated in the context of realistic benchmark particle physics applications.\\
%In particular, the processing speed of ML techniques for inference is the major determining factor for  applications like real time processing.

If different ML techniques can achieve equivalent physics performance but require different processing power, it is important to quantify what is driving the performance gains and what level of performance-processing power trade-off is maintainable to achieve the required physics goals.

\subsection{Graphical Processing Units}\label{subsec:GPU}

GPUs have been extremely useful in speeding up the training of complex deep learning models. Many researchers in HEP are currently relying on private or university GPU clusters to perform machine learning training. Unfortunately, there is no centralized GPU resource available for general HEP usage. Availability of greater and more modern GPU resources would significantly reduce the training time and assist many on-going $R\&D$ efforts in the community.

\subsection{Cloud TPUs}\label{subsec:TPU}

Cloud TPUs systems are built around Google's TPUs, which are custom silicon chips for machine learning.
Cloud TPU Pods are multi-rack supercomputers that can include more than 1,000 TPU chips as well as many host machines.
Cloud TPUs and Cloud TPU Pods are optimized to accelerate the dense linear algebra that is commonly found in cutting-edge deep neural network models, but they can potentially be used for other purposes as well.
Current efforts in the HEP community are studying the use of Cloud TPUs for possible acceleration of the training stage.
Cloud TPUs and Cloud TPU Pods are available to the public via Google Cloud~\cite{cloudgoogle}.

\subsection{High Performance Computing}

Resource-rich many-core processors such as MIC, GPUs, and TPUs are vital to the optimization of the training time of most modern machine learning algorithms, including deep neural networks, generative adversarial networks, autoencoders, etc. Availability of High Performance Computing (HPC) resources equipped with many-core processors and high-performance network storage are essential to distributed running of large-scale machine learning algorithms. Current efforts to bring and expand the availability of HPC resources in high-energy physics computing will be vital to the successful progress of the application of machine learning techniques to current and future experiments.

%and help address the following physics needs: object and event reconstruction, real-time analysis, fast simulation and event analysis. %Initial exploratory studies[\textbf{reference?}] have been conducted on running ML applications for HEP and are promising.

\subsection{Field Programmable Gate Arrays}
Field Programmable Gate Arrays (FPGAs) provide an efficient and low-latency application of machine learning algorithms directly at the level of hardware, as desirable for HEP trigger systems. The following ML algorithms are more suitable for FPGAs due to their simpler parallelization: boosted decision trees, random forests and decision rule ensembles. For example the CMS experiment currently uses boosted decision trees in FPGAs in the trigger system to estimate muon momenta. Further research and development is needed in this area to apply more advanced machine learning techniques like deep learning directly in the hardware. One of the challenges is the limited availability of floating point operations gates and the precision needed to maintain the best performance. The possibility of coupling the FPGAs with a CPU with significant random-access memory (RAM) allows the shift of some of these operations to RAM.

%In particular, it is foreseen that the advent of more libraries implementing an abstraction layer to neural nets inference on such hardware will greatly help the adoption and development of deep lerning technology.
%The use of such hardware is likely more taylored for inference than for training.

\subsection{Opportunistic Resources}
The current HEP computing model is based on a tiered structure where computing resources are mostly large data centers providing CPU resources for collaboration. Although existing resources are gradually moving towards supporting GPUs, it is unlikely to reach all HEP computing centers in the near future. Therefore opportunistic resources are a possible option for training machine learning applications.\\

%Below is a list of available resources for
%\begin{itemize}
%   \item lxplus, lxbatch, analytix clusters at CERN or lpc at FNAL
%   \item Tier 1 and Tier 2 centers
%   \item HPC centers
%   \item Experiment trigger farms during accelerator downtime
%   \item Cloud providers
%\end{itemize}

Currently, cloud solutions provided by the industry run ML workflows on dedicated hardware and offer interfaces for training machine learning models. The scientific community should work closely with cloud providers to harmonize our analysis computing needs and data access patterns with their business models. Costs of the cloud resources should be compared with the costs of procuring these resources independently.\\

In order to make the best use of resources available to the community, all resources should ideally be made available through a unique work queue. That implies some uniformization of the software stack, and several specific requirements in the resource management system, especially in terms of data movement.

\subsection{Data Storage and Availability}
Data storage limitations will have a major impact on machine learning applications. Presently, to train machine learning algorithms, it has been possible to take advantage of increases in statistics of Monte-Carlo simulated events needed for other use cases. Further machine learning progress may require more simulated data than what is available today. How to produce and store these additional large amounts of data is a challenge that needs to be overcome.\\

Availability of data at PByte/EByte scale represents another challenge for the ML community.
A good solution must provide access to a large data volume for hundred or thousand of users simultaneously.
Additionally, data movement might need to be automatized to make the training data available transparently at high speed local storage with use of an automatic caching mechanism.
The success of Apache Spark and Google BigQuery platforms may serve as a model.
In addition to the regular HEP workflows, data streaming, transformation and readout in mini-batches may be required to train models over large data sets.

\subsection{Software Distribution and Deployment}
To efficiently use the resources described in previous subsections, machine learning software needs to be available on the computing resources. Platforms, such as the CERNVM File System (cvmfs), are very useful for distributing software instead of requiring local installation. Additional tools like docker containers for application shipping can be useful in providing homogeneous software environments across the different systems. Another challenge is that the software layer needs to be agnostic to the hardware back-end.
% %Issues are sometimes with the needs for synchronization of various versions of software, point at which the use of self-consistent container is indeed more practical.
%At the same time we need tools to discover these resources and have clear understanding pricing model of commercial solutions.

\subsection{Machine Learning As a Service}
Current cloud providers rely on the machine learning as a service model, allowing for the efficient use of common resources, and make use of interactive machine learning tools. Machine learning as a service is not yet widely used in HEP, but examples of successful publications which used machine learning as a service exist, e.g.~\cite{Aaij:2014azz}. Specialized HEP services for interactive analysis, such as CERN's Service for Web-based Analysis (SWAN) may play an important role in adoption of machine learning tools in HEP workflows. In order to use these tools more efficiently, sufficient and appropriately tailored hardware resources  described in this chapter are needed.
In addition, the evolution of current HEP reconstruction software model to support off-loading ML tasks to heterogeneous computing resources, as well as ML task pooling, should be studied.

\section{Training the community}
\label{sec:training}
The CWP on training~\cite{HSF-CWP-2017-02} discusses the training needs of the HEP community in detail.
In order to address the communication barrier and to speak the same language, training in machine learning concepts and terminology have to be part of a standard curriculum.
%The training should focus on well-maintained and well-documented software packages.
In addition to lectures also hands-on tutorials on specific tools are necessary.
Being able to apply machine learning to practical HEP problems requires the understanding of basic machine learning concepts and algorithms.
For this, regular data science lecture series and seminars, like~\cite{mlhep}, are very useful.
Also university level courses dedicated to machine learning applications in physics research are an excellent way to train undergraduate and graduate students.
% For example, the ``Deep Learning in Physics Research'' course with 60 participants consisting of 12 lectures and exercises which are performed on 20 GPUs of the VISPA internet platform~\cite{vispa}. %Such resources should be made more centrally available.
%
Experimental collaborations currently have training activities for newcomers that focus on analysis software and introduction to domain knowledge~\cite{2016chep.confE.334B}.
Machine learning should next be incorporated into the incoming collaborators' training efforts of the experiments.
%As discussed in the Community White Paper report on Careers and Training~\cite{HSF-CWP-2017-02}, ensuring the development and availability of resources for knowledge transfer is likewise essential to ML.

\section{Roadmap}
\label{sec:roadmap}
% ---------------------------------------------------
% CHAPTER SUMMARY MESSAGE: < WRITE THIS BEFORE ANY EDITING>
% ---------------------------------------------------

In this chapter we discuss the roadmap towards implementation of the research and development areas described in Section~\ref{sec:applications}.

\subsection{Timeline}
The incorporation of machine learning into particle physics experiments must respect two primary time lines: the schedules of HL-LHC and funding agencies, and the experiments' need for extensive validation of the algorithms.\\

The current LHC schedule has Run 3 starting in 2021 and the HL-LHC starting in 2026. As software processes and algorithms are re-imagined, their implementation must fit into these time frames if they are to maximize their benefit to the particle physics community. To fit this schedule, a newly proposed implementation would need to have a demonstration in 2018 to prove viability. Two years later, in 2020, the proposed idea needs to attain a level of maturity to be included in the HL-LHC Technical Design Report. The project should then be further refined towards a large scale test around the middle of Run 3, about 2022. Run 3 is scheduled to end in late 2023, after which point the project must then be adapted to the HL-LHC software and physics analysis environment as it will be required by the experiment.
%The steps necessary to incorporate new tools into existing analysis implementations are driven %primarily by the experimental validation process. as well as the two timescales mentioned above.% The point is this is research and development roadmap,
%

\subsection{Steps to Deployment}
The path of taking a machine learning idea from conception to community-wide acceptance and deployment will entail several stages, as appropriate. There are ample opportunities to make the process more efficient.

\begin{enumerate}
 \item \textbf{Problem formulation and data set preparation}: Problem formulation is the first step in building a machine learning algorithm. The inputs and desired output need to be established. The training and validation data sets must be identified and simulated. In many cases, these data sets are large, and resources must be identified to possibly create and store the data. In most cases, the data needs to be processed into a form suitable for input into the algorithm. As discussed in Section~\ref{subsec:benchmark}, common benchmark samples will later on facilitate the comparison of different approaches.
 \item \textbf{Feasibility and demonstration}: Given a dataset, appropriate machine learning algorithms need to be investigated and evaluated for their suitability to solve the problem.
 \item \textbf{First application}: An application of the solution to one or a few specific physics analysis examples where the ML technique significantly improves the result. The incorporation of the technique into the computing work-flow will likely be very specific to the application and require significant manual intervention.
 \item \textbf{Scaling and optimization}: Evolving from a demonstration to a general solution requires the use of realistic data sets with full detector simulation. Furthermore, the solution will also require optimization to achieve nominal physics and computing performance. This stage will likely require significant computing resources to scale solutions to the full detector and data sets.
 \item \textbf{Integration and Validation}: The solution needs to be incorporated into the experimental software and work-flow and must be validated.
\end{enumerate}

As an example, an effort has recently started to build generative models that can significantly accelerate the simulation of particle showers in calorimeters.
These early efforts are based on simplified data sets specifically created for this problem, without the complications of realistic data and limited to a small calorimeter sections.
The first papers~\cite{Paganini:2017hrr} use generative adversarial networks to generate calorimetric data which are reasonably faithful but still require tuning. The next step involves exploration of neural network architectures and systematic hyper-parameter scans to achieve the required performance.\\
% INCONCLUSIVE. Either they require full sim, or they don't: The technique can be applied to several possible searches at LHC that involve boosted objects, where simulation samples require full GEANT-based simulation and are therefore limited in statistics due to resource limitations.

The process of employing the new technique in a publication will elicit scrutiny by the full experiment, effectively validating the technique. Once the technique is accepted, it can be generalized beyond this first application and then incorporated into the experiment's software for use by others. Finally, as the technique is applied to an increasing number of physics analyses, the technique will be incorporated into the experiment's production work-flows.

\section{Conclusions}
\label{sec:conclusions}
% Conclusions
% too generic, and too specific at the same time. applies to all CWPs, but doesn't cover nearly all applications from the earlier parts of the CWP: As particle physics moves into the post-Higgs boson discovery era, the physics drivers of the High-Luminosity Large Hadron Collider and future neutrino experiments will require increasingly more powerful identification and reconstruction algorithms to extract rare signals from copious and challenging backgrounds.

Machine learning algorithms are already state of the art in many areas of particle physics and will likely be called on to take on a greater role in solving upcoming data analysis and event reconstruction challenges.
In this document we have outlined the promising areas of research and development applications of machine learning in particle physics and focused on addressing the most important science drivers.
We identified the need for greater collaboration with external communities in machine learning and a need to train the particle physics community in machine learning.
We also identified how these prerequisites for successful incorporation of machine learning applications into high-energy physics  can be met and provided an example roadmap for the implementation of machine learning applications into the workflows of particle physics experiments.

\section{Acknowledgements}
\label{sec:acknowledgements}
Vladimir Vava Gligorov acknowledges funding from the European Research Council (ERC) under the European Union's Horizon 2020 research and innovation programme under grant agreement No 724777 ``RECEPT''.\\
Omar Zapata is supported by Sostenibilidad-UdeA and COLCIENCIAS through the grant No 111577657253.

\clearpage

\appendix
\section{Appendix}
\label{sec:appendix}
\subsection{Matrix Element Methods}
\label{subsec:MEM}

The ME method is based on \emph{ab initio} calculations of the probability density function $\mathcal{P}$ of an event with observed final-state particle momenta ${\bf x}$ to be due to a physics process $\xi$ with theory parameters $\boldsymbol\alpha$.
One can compute $\mathcal{P}_{\xi}({\bf x}|{\boldsymbol\alpha})$ by means of the factorization theorem from the corresponding partonic cross-sections of the hard-scattering process involving parton momenta ${\bf y}$ and is given by
\begin{equation}
 \mathcal{P}_{\xi}({\bf x}|{\boldsymbol\alpha}) = \frac{1}{\sigma^{\rm fiducial}_{\xi}(\boldsymbol\alpha)} \int d\Phi ({\bf y}_{\rm final}) \; dx_1 \; dx_2~\frac{f(x_1)f(x_2)}{2s x_1 x_2} \; |\mathcal{M}_{\xi}({\bf y}|\boldsymbol\alpha)|^2 \; \delta^{4}({\bf y}_{\rm initial}-{\bf y}_{\rm final}) \; W({\bf x}, {\bf y})
 \label{eqn:MEProb}
\end{equation}
where and $x_i$ and ${\bf y}_{{\rm initial}}$ are related by $y_{{\rm initial},i}\equiv \frac{\sqrt{s}}{2}(x_i,0,0,\pm x_i)$, $f(x_i)$ are the parton distribution functions, $\sqrt{s}$ is the collider center-of-mass energy, $\sigma^{\textrm{ fiducial}}_{\xi}(\boldsymbol\alpha)$ is the total cross section for the process $\xi$ (with $\boldsymbol\alpha$) times the detector acceptance, $d\Phi({\bf y})$ is the phase space density factor, $\mathcal{M}_{\xi}({\bf y}|\boldsymbol\alpha)$ is the matrix element (typically at leading-order (LO)), and $W({\bf x}, {\bf y})$ is the probability density (aka ``transfer function'') that a selected event ${\bf y}$ ends up as a measured event ${\bf x}$.
One can use calculations of Eq.~\ref{eqn:MEProb} in a number of ways (e.g. likelihood functions) to search for new phenomena at particle colliders.

%%%

As stated in Sect.~\ref{sec:applications-MEM}, the ME method has three notable features: it (1) does not require training data being an \emph{ab initio} calculation of event probabilities, (2) incorporates all available kinematic information of a hypothesized process, including all correlations, and (3) has a clear physical meaning in terms of the transition probabilities within the framework of quantum field theory.\\

In reference to point (1), the matrix element $\mathcal{M}_{\xi}({\bf y}|\boldsymbol\alpha)$ in the method involves all partons in the $n\rightarrow m$ process, so when the 4-momentum of particles are not completely measured experimentally (e.g. neutrinos), one must integrate over the missing information which increases the dimensionality of the integration.
In reference to point (2), a clever technique to re-map the phase space in order to reduce the sharpness of integrate in that space in an automated way ({\sf MADWEIGHT}~\cite{Artoisenet:2010cn}) is often used in conjunction with a matrix element calculation package ({\sf MADGRAPH\_aMC\@NLO}~\cite{Alwall:2014hca}).
In practice, evaluation of definite integrals by the ME approach invokes techniques such as importance sampling (see {\sf VEGAS}~\cite{PETERLEPAGE1978192,Ohl:1998jn} and {\sf FOAM}~\cite{JADACH200355}) or recursive stratified sampling (see MISER~\cite{Press:1989vk}) Monte Carlo integration.
Acceleration of some of these techniques on modern computing architectures has been achieved, for example concurrent phase space sampling in VEGAS on GPUs.

\clearpage

\printbibliography[title={References}]

\end{document}